\documentclass[%
 aip,
 amsmath,amssymb,
 reprint,%
]{revtex4-1}

\usepackage{graphicx}% Include figure files
\usepackage{dcolumn}% Align table columns on decimal point
\usepackage{bm}% bold math
\usepackage[utf8]{inputenc}
\usepackage[T1]{fontenc}
\usepackage{mathptmx}
\usepackage{subfigure}

\begin{document}

\preprint{AIP/123-QED}

\title[Ventilation study of a public washroom]
{Effect of recirculation zones on the ventilation of a public washroom}
% {Air flow management in a public washroom using simple flow visualization experiment}
% Force line breaks with \\

\author{Krishnendu Sinha}
 \email{krish@aero.iitb.ac.in} 
% \homepage{http://www.Second.institution.edu/~Charlie.Author.}
\affiliation{Indian Institute of Technology Bombay, Mumbai, India}

\author{Utkarsh Verma}
\affiliation{Indian Institute of Technology Bombay, Mumbai, India}
\author{Janani Srree Murallidharan}
\affiliation{Indian Institute of Technology Bombay, Mumbai, India}
\author{Vivek Kumar}%
\affiliation{Ansys India Pvt. Ltd., Pune, India}%

% {\it Indian Institute of Technology Bombay, Mumbai, India}

\date{\today}% It is always \today, today,

%\if{false}
\begin{abstract}

Air-borne transmission can pose a major risk of infection spread in enclosed spaces. Venting the air out using exhaust fans and ducts is a common approach to mitigate the risk.
In this work, we study the air flow set up by an exhaust fan in a typical shared washroom that can be a potential hot spot for COVID-19 transmission. The primary focus is on the regions of recirculating flow that can harbor infectious aerosol for much longer than the well-ventilated parts of the room.
Computational fluid dynamics is used to obtain the steady state air flow field, and Lagrangian tracking of particles give the spatial and temporal distribution of infectious aerosol in the domain. 
It is found that the washbasin located next to the door is in a prominent recirculation zone, and particles injected in this region take much longer to be evacuated. The ventilation rate is found to be governed by the air residence time in the recirculation zone, and it is much higher than the time scale based on fully-mixed reactor model of the room. Increasing the fan flow rate can reduce the ventilation time, but cannot eliminate the recirculation zones in the washroom.

\end{abstract}
%\fi 
\maketitle

\section{Introduction}

% Air-borne transmission
Air borne transmission is a prominent concern for highly infectious diseases like COVID-19.
Tiny droplets and aerosol can be carried in the air flow over a distance, potentially exposing a large number of people to the infection.
This is particularly dangerous in indoor spaces, where the infectious aerosol can be present in the air for a long duration.
% [[ Knowing the air flow, especially in enclosed spaces, and its management becomes important to mitigate the risk. \cite{} ]]
Ventilation of public spaces and shared facilities is therefore an active area of study in the context of Covid-19. \cite{mathai_science_2021,walking_pof_2020,classroom_pof_2020,classroom_pof_2021}
It is recommended that the air in enclosed spaces is replaced with fresh air at regular intervals.
Ventilation and air conditioning systems are rated for number of air changes per hour (ACH), which is based on the volume of the enclosed space and the air flow rate through the ducts and fans. \cite{ventilation_jfm_2020}
%Typical guidelines recommend ACH higher than 6-12 depending on the nature of the shared space.
% It assumes that the air in the room is fully mixed and that the infectious aerosol is uniformly distributed in the space. 
This gives an estimate of the average residence time or the mean age of air in the room.

% Air flow pattern (recirculation zone)
The actual residence time of air in an enclosed space and the related probability of air-borne transmission depends on the air flow pattern set up by the ventilation system.\cite{classroom_pof_2021}
Specifically, the location of ducts and vents, relative to the geometry of the room play a key role.\cite{Train_bldsim_2014,airplane_ast_2020}
The inlet and outlet ports set up an air circulation pattern in a given room, which determines the pattern of aerosol spread from a potential source. \cite{boxcleaner_pof_2021}
Of particular interest are the recirculation zones formed at the corners of a room, and around obstacles.\cite{classroom_pof_2020, musicroom_pof_2021}
Such regions of recirculating flow are characterized by low air flow velocity and high residence time.
Naturally, infections particles can remain in these pockets, while the other parts of the room are well ventilated. The deposition of aerosol on surfaces is also found to be correlated to the location and size of recirculation zones. \cite{boxcleaner_pof_2021}

% The placement of mitigation measures like air purifiers can also influence the air flow pattern.
% The infection spread is dependant on the air flow pattern around a potential source. \cite{drikakis_pof_2020}

\if{false}
A common practise is to have cross-ventilation, where fresh air enters the room from one side, and exits from another side. This sets up a primary flow of air between the entry and exit ports, as desired for quick ventilation.
However, pockets of secondary flow or recirculation regions are usually present at the corners of a room, and around obstacles.\cite{classroom_pof_2020, musicroom_pof_2021}
% At times, the secondary flow can cover a large fraction of the enclosed space, undermining the usefulness of the air flow  management system. \cite{hospital_jcp_2021}
Such regions of recirculating flow are characterized by low air flow velocity and high residence time.
In rooms with multiple entry and exit ports distributed around the room, the air flow pattern can be more complex, and can greatly influence the infection spread and ventilation of the room.

% Identifying such regions and designing ducts and vents to minimize and possibly eliminate them is therefore important.  {\bf Do we need to tone this down a bit?}

\fi

% CFD, airflow, mitigation measures
Computational fluid dynamics is a powerful tool to study the air flow pattern in enclosed spaces. In the context of Covid-19, the objective is to analyze how the air flow carries the infectious droplets and aerosol from a potential source to susceptible individuals in the room.\cite{classroom_pof_2021}
The concentration of droplets in a given region over a period of time is used to assess the probability of infection spread. % It can give an estimate of how long it is safe to use a
Several mitigation strategies have also been evaluated, including air filters, air purifiers, opening doors and windows, and enhancing the flow rate of the airconditioning system.
It is found that the efficacy of these mitigation measures depend critically on the air flow pattern in the room and the location of the source of infection.\cite{musicroom_pof_2021,boxcleaner_pof_2021,restaurant_pof_2021}

There are several studies of the air flow in enclosed spaces. These include passenger vehicles,\cite{mathai_science_2021} classroom,\cite{classroom_pof_2020,classroom_pof_2021}
restaurant,\cite{restaurant_pof_2021,Cafeteria_pof_2021} health care facilities,\cite{hospital_jcp_2021} elevators\cite{Elevator_pof_2021} and public transport.\cite{Train_bldsim_2014,Bus_pof_2021}
It is clear that the air flow and aerosol distribution pattern varies from one configuration to another, and with changes in the location of source.
Here, we study the air flow pattern and aerosol spread in a washroom set up, with a focus on recirculation zones and their effect on the rate of evacuation of infectious aerosol.
% To the best of our knowledge, there has been no study of air flow pattern in a washroom setup.
Washrooms are usually shared facilities in offices, schools, restaurants and other public spaces, with a large potential of infection spread. Active use of water in toilet flushing and the wash basin can be a major source of droplets. These droplets can rise high into the air,\cite{toilet_pof_2020,urinal_pof_2020} and some of them can remain in the air for significant duration of time.

\begin{figure*}[t!]
{\includegraphics[width = 0.95\textwidth,trim={0cm 0.0cm 0cm 0cm },clip]{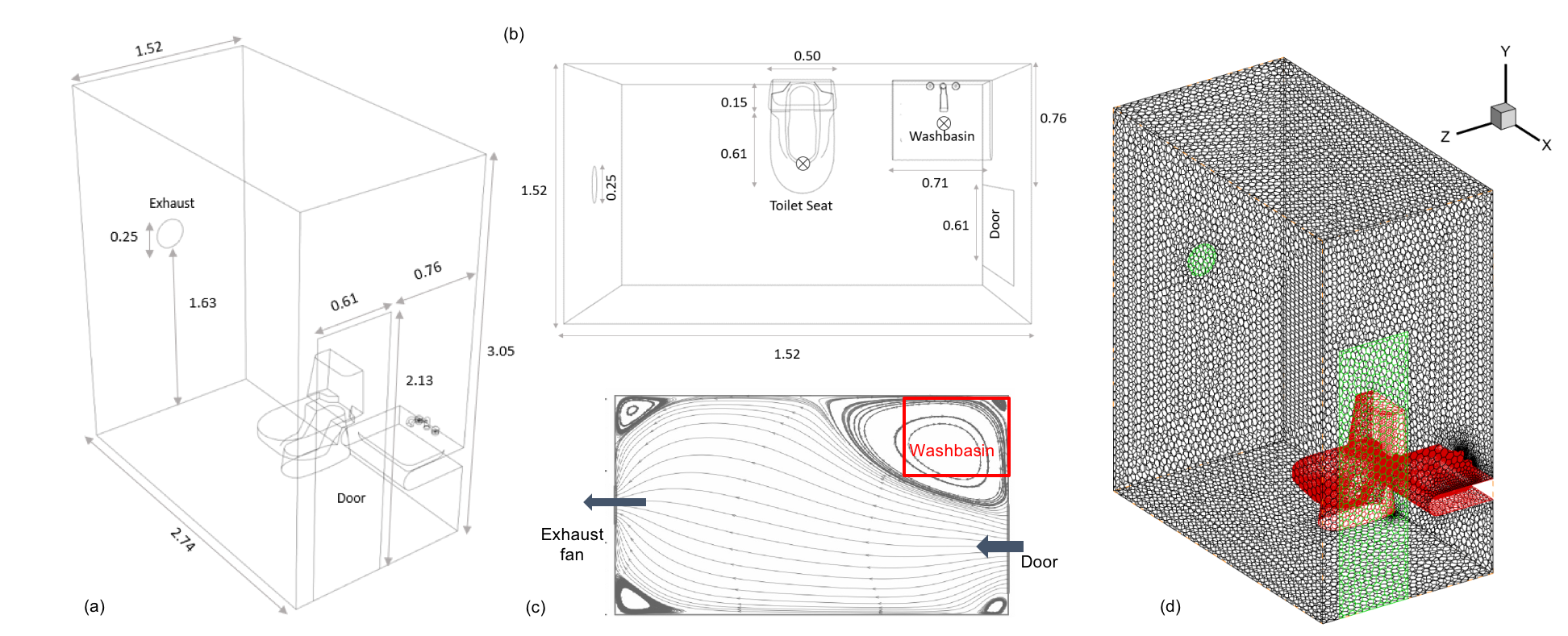}} 
\caption{A typical washroom geometry showing an open door, exhaust fan, a washbasin and a toilet seat (a) isometric view, (b) top view (c) a typical air flow pattern in top view and (d) computational mesh for 3D simulation.}
\label{geom} 
\end{figure*}

A typical single-person washroom geometry is shown in Fig.~\ref{geom}, with a wash basin next to the door and a toilet seat located in the center of the washroom. Such washrooms are used by multiple people, one after another, and are commonly found in densely-populated areas in India. To reduce infection spread from one user to another, it is recommended that the washroom be sanitized after every use. This can be achieved by disinfectant spray and UV-based sanitization. A common practice is to have an exhaust fan evacuate the air out of the washroom. It is recommended that the door be kept open between two consecutive usages of the washroom to increase ventilation.
A closed door can hamper the efficacy of the exhaust fan significantly, and the volume flow rate through the fan can be substantially reduced.

% The washroom is 9 x 5 x 10 cu. ft., and other dimensions, including the door, are marked in Fig. 1. A 25 cm dia exhaust fan is installed on the wall directly opposite to the door for venting out the air from the  room.
The volume flow rate of a typical exhaust fan used in washrooms is 270 m$^3$/h, or 158.8 cubic feet per minute (CFM). For the dimensions of the washroom shown in Fig.~\ref{geom}, the fan can vent out an equivalent volume of air in about 170 s. This gives 21.2 air changes per hour (ACH), which means that the fan will refresh the air in the washroom 21.2 times in an hour, or every 170 s. It is based on the assumption that the air in all parts of the room are vented at the same rate. This is not the case in reality.
The ACH timescale is also inherent in the mixed-reactor model of ventilation,\cite{indoor_PNAS_2021} which  assumes that the infectious aerosol is uniformly distributed in the volume of air in the room.
CFD simulations in a variety of indoor spaces \cite{restaurant_pof_2021, musicroom_pof_2021} show that the aerosol distribution critically depends on the air circulation pattern in the room.
Of particular interest are the pockets of air in recirculation zones that can be trapped for much longer than the ACH timescale for a fully-mixed reactor model.

The objective of this paper is to use CFD to study the air flow pattern in the washroom, and how it affects the rate at which infectious particles are vented out by the exhaust fan.
We specifically identify the regions of recirculating air over the washbasin and near the toilet seat, which can be particularly important in terms of infection spread. The residence time of air in these recirculation zones is compared with that in the primary flow setup between the open door (inlet) and the exhaust fan (exit).
We inject particles (simulated droplets and aerosol) in different part of the washroom and use Lagrangian tracking to quantify their spatial spread in the domain. We also study the time history of their evacuation from the washroom, and compare it with the ACH ventilation rate given by the fully-mixed reactor model. 
Finally, we evaluate the effect of increasing the exhaust fan CFM to vent the particles out of the domain faster, and whether it can eliminate the recirculation zone over the washbasin.
% Finally, we evaluate mitigation strategies in terms of increasing the CFM of the exhaust fan, and placing additional fans to increase the turbulence level in the room.

% Section 2 presents the computational methodology, both for the air flow simulation as well as the Lagrangian particle tracking. This is followed by results in section 3 that describes the air flow pattern in terms of the reciruclation zones observed in the washroom. Conclusions are presented at the end.

% \vskip3cm

%%%%%%%%%%%%%%%%%%%%%%%%%%%%%%%%%%%%%%%%%%%%%%%%%%%%%%%

%%%%%%%%%%%%%%%%%%%%%%%%%%%%%%%%%%%%%%%%%%%%%%%%%%%%%%%

%\clearpage
\section{Simulation Methodology}

% Solver details
We solve the Reynolds-averaged Navier Stokes (RANS) equations for incompressible flows using commercial software ANSYS Fluent (2020 R2 version) \cite{ansys}. 
SIMPLE algorithm is used for pressure-velocity coupling, and the realizable $k-\epsilon$ model \cite{Shih} is used for turbulence closure. 
A second-order upwind scheme is used for the convective terms of the RANS equations, whereas a first-order method is employed for the turbulence transport equations. 
Iterative convergence is achieved when the scaled residual drops below 10$^{-5}$ for the mean and turbulence variables.

% Grid details
Unstructured polyhedral mesh is shown in Fig.~\ref{geom}d, where the exhaust fan and the door are modeled as inlet and outlet boundaries respectively.
The washbasin and other fixtures are included in the mesh, and dimensions are given in Fig.~\ref{geom}a and \ref{geom}b. 
A human is not modeled in the geometry in line with the problem definition, where the ventilation of the washroom is studied between two usages.
The mesh consists of $3.2 \times 10^5$ elements, with a grid resolution of 0.05 m in the interior of the domain. The mesh is much finer near solid boundaries, and the wall-normal spacing corresponds to $y^+ \simeq 10$ or lower along the walls. The values are relatively higher near the exhaust fan, but it is not expected to affect the rest of the flow field.
The flow field results are relatively insensitive to the grid resolution in this range; a coarser grid with 0.1 m interior cell size gives almost identical results for the air flow pattern in the washroom. The variation in the velocity magnitude is found to be within 1\% between the two grids.
% A grid refinement study using grid resolution of 0.10 and 0.01 m give comparable air flow pattern in the washroom. The velocity magnitude difference between the different grid is within ???\% 

\begin{figure*}[t!]
{\includegraphics[width = 1.05\textwidth,trim={0cm 0.0cm 0cm 0cm },clip]{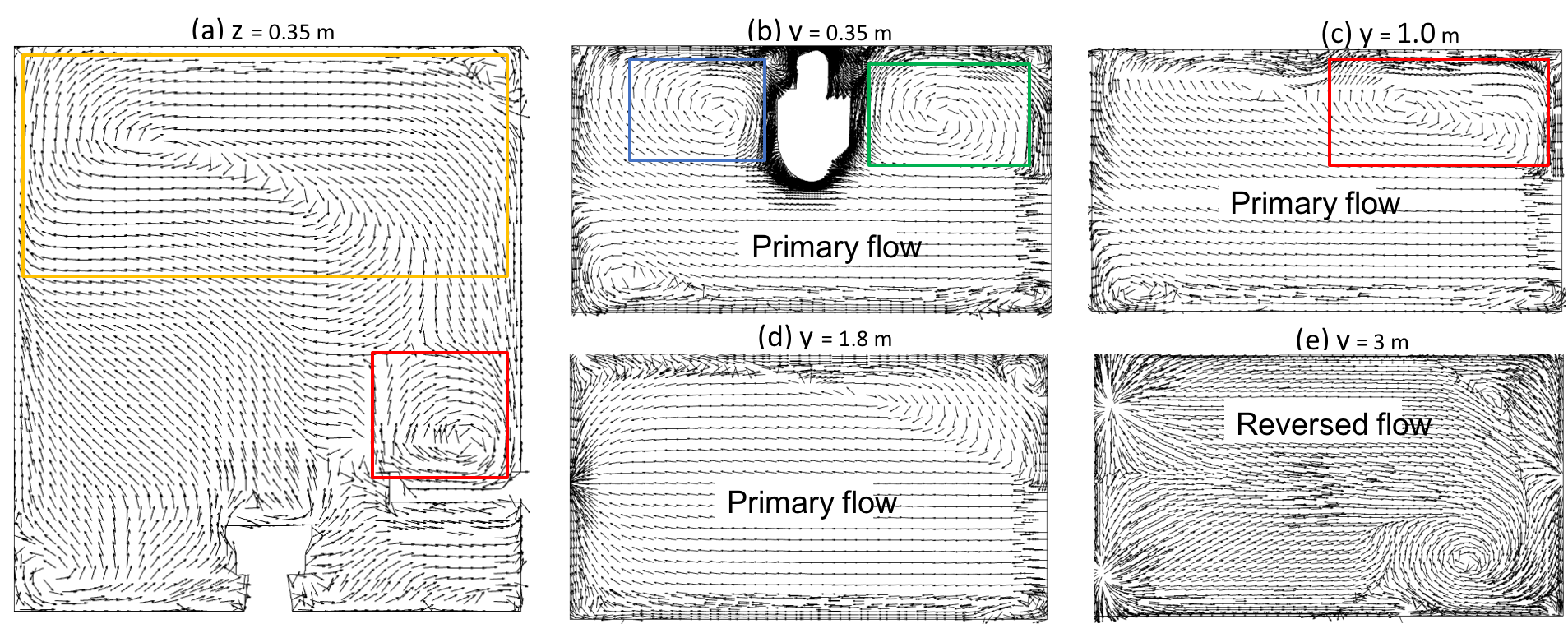}} 
\caption{Velocity vectors at different horizontal (top view) and vertical (side view) planes to visualize the three-dimensional flowfield computed in the washroom. The recirculating regions over the washbasin (red box), under the washbasin (green box), next to the toilet seat (blue box) and near the ceiling (yellow box) are identified in the plots.}
\label{vel_vect} 
\end{figure*}

% Boundary conditions
The fan CFM is prescribed as an equivalent velocity boundary condition ($V_{fan}$ = 1.5 m/s) at the exit and the door is specified as a pressure inlet boundary with zero gauge pressure.
Rest of the boundaries, including the floor, the ceiling, as well as the surfaces of the washbasin and toilet seat, are prescribed as no-slip viscous wall.
Turbulence intensity of 5$\%$ is used at the inlet and exit boundaries, along with enhanced wall treatment for the $k-\epsilon$ equations at the no-slip walls.\cite{Wall}
The inlet value of the turbulent viscosity is prescribed as ten times the dynamic viscosity, which is evaluated at a temperature of 300 K using Sutherland's law.
Similar methodology has been applied to air flow simulation in enclosed spaces.\cite{classroom_pof_2020}

% {\bf Add a couple paragraph on particle tracking simulation methodology}

The steady state air-flow solution obtained in the domain is used to perform Lagrangian particle tracking, as per the equation
$$ m\frac{d \vec v}{dt} = m \vec g + \vec F_L + \vec F_D $$

\noindent
where $m$ and $\vec v$ are the mass and velocity of a particle and $\vec g$ is the acceleration due to gravity. 
The drag force $\vec F_D$ is given by the spherical drag model and $\vec F_L$ is Saffman's lift force.\cite{classroom_pof_2020} Additional effects due to pressure force and virtual mass force are neglected owing to the small size of the particles. On the other hand, the particles are assumed to be large enough to neglect Brownian forces.
The effect of turbulent dispersion on particle trajectories is modeled using discrete random walk method\cite{ansys} and the procedure is similar to that in recent studies. \cite{classroom_pof_2020}

% \if{false}
 
The particles are taken as water droplets of 1 micron diameter. A mono-dispersed collection is used to study the effect of air flow on particle dynamics. The particle diameter is varied subsequently to study its effect on the ventilation rate. 
Two injection locations are considered, namely, over the washbasin and on top of the toilet seat. These are frequently-used location, with significant water usage. Toilet flushing can be an important generator of droplets and aerosols.\cite{toilet_pof_2020}
A total of $3.5 \times 10^5$ particles are injected at one of these locations at a height of 1 m, with a nominal injection velocity of 0.1 m/s. The total particle mass adds up to $2.85 \times 10^{-9}$ kg that is injected over 0.5 s. The particles are assumed to escape the domain, once they hit a solid wall. 
The particles may stick to the wall with film formation. However, since our objective is limited to airborne transmission, and not the contamination of surfaces, we have opted for escape boundary condition in the majority of the simulations presented in the paper. 
The effect of other boundary conditions, in the form of particles getting trapped or reflected from the wall are considered at the end.

Evaporation is not modelled explicitly, given the high variability of evaporation rates with relative humidity and temperature conditions prevailing across different geographical locations. Note that the majority of washrooms in India do not have any form of heating or air-conditioning. In addition, the relative humidity in a washroom with shower facilities can be very high, thus slowing down the evaporation significantly. Finally, a droplet may not evaporate completely and the salt nuclei left behind can be potentially carry the virus in the air for much longer than typical evaporation time scales. A qualitative assessment of the effect of evaporation can be deduced from the effect of varying the droplet diameter, such that a larger droplet will evaporate to a smaller diameter. The results presented in this study are found to largely insensitive to the changes in droplet diameter. It points to the fact that the effect of recirculation regions on the ventilation rate holds irrespective of whether evaporation rates are accounted for or not.

A fluid residence time is also calculated based on
a user-defined scalar in ANSYS Fluent. A differential equation is solved for the flow time, obtained as the ratio of the local grid size and velocity. The equation is integrated over the steady state velocity field, with a zero value specified at the inlet boundary. The value of the flow residence time thus obtained gives the time taken for a fluid parcel to traverse the distance along a flow streamline. The higher the value, the longer it takes for a parcel of fresh air to reach a given point from the door inlet. Conversely, low residence time is associated with quick ventilation. A distribution of fluid residence time is used to identify the well-ventilated regions of the washroom, as opposed to the regions of trapped air in recirculation zones or dead-air regions.

\begin{figure*}[t!]
{\includegraphics[width = 1.0\textwidth,trim={0.cm 0.0cm 0cm 0cm },clip]{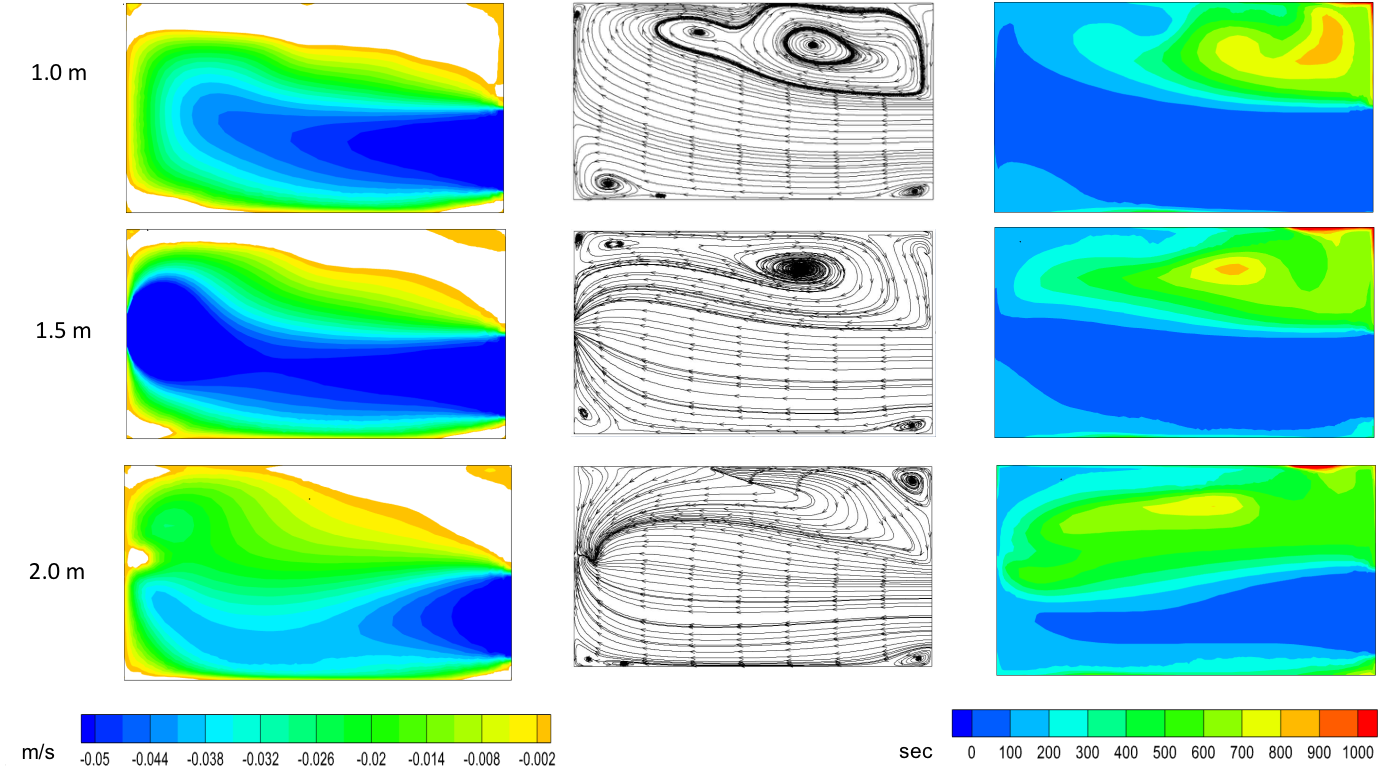}} 
\caption{Contours of x-velocity (left), streamlines (center) and air residence time (right) plotted in different horizontal sections of the washroom: $y$ = 1 m (top), $y$ = 1.5 m (middle) and $y$ = 2 m (bottom). The recirculation zones are marked by reversed flow (white regions) and high flow residence time.}
\label{flowfield} 
\end{figure*}

%%%%%%%%%%%%%%%%%%%%%%%%%%%%%%%%%%%%%%%%%%%%%%%%%%%%%%%

%%%%%%%%%%%%%%%%%%%%%%%%%%%%%%%%%%%%%%%%%%%%%%%%%%%%%%%

%\clearpage
\section{Results}

\subsection{Air flow simulation results}

Figure \ref{vel_vect} shows the computed flowfield solution in terms of velocity vectors in one vertical and several horizontal cross-sections. The vertical plane at $z$ = 0.35 m passes through the washbasin and the toilet seat. The horizontal plane at $y$ = 0.35 m is located below the washbasin, $y$ = 1.0 m plane is above the washbasin level, $y$ = 1.8 m passes through the exhaust fan and $y$ = 3.0 m is close to the ceiling. 
%%%
A primary flow is set up (from right to left) between the door and the exhaust fan, and we see uniform velocity vectors in all the horizontal planes, except for the one near the ceiling. Recirculation regions can be observed either at the corners of the washroom, for example, over and under the washbasin, or around obstacles like the toilet seat. There is a large region of reversed flow (left to right) near the ceiling (at $y$ = 3.0 m), and it is part of the recirculating flow visible in the top part of the vertical plane. We are particularly interested in the recirculation zone formed over the washbasin (marked by a red box) and its role in trapping infectious particles.

Additional details of the three-dimensional flow field are shown in Fig.~\ref{flowfield} in terms of $x$-component of velocity, streamlines and residence time of air. Once again, the data is plotted in three horizontal planes at heights of 1.0 m, 1.5 m and 2.0 m. The residence time of air is calculated based on the procedure described in section 2 and it gives the mean age of air in different parts of the room. Low residence time in a region implies that it is well-ventilated, while long residence time of air would indicate dead-air zones. The residence time of air also provides a time-scale of venting infectious particles from the domain, as discussed subsequently.

The blue region in the velocity contour plots (left panel in Fig.~\ref{flowfield}) corresponds to the primary flow originating from the door. The magnitude of velocity is high in this region, and it is directed towards the negative $x$-axis (from door inlet to fan exit location).
The streamlines in the primary flow indicate a direct path from the inlet to the outlet station, and the flow residence time in this region is relatively low (about 50 s). 
%%%
A quick order of magnitude estimate of the time required for air to reach from the door to the exhaust fan can be made as follows. Applying mass conservation between the inlet and outlet locations gives us an approximate air velocity of 5.7 cm/s at the door. This can be assumed to be the characteristic velocity of the primary flow. Taking a ratio of the velocity with the length of the washroom gives the primary flow timescale $\tau_P \simeq $ 50 s.
This means that the primary air stream entering the washroom through the door will exit through the fan within this time scale. The primary flow region is thus considered well ventilated.

%\begin{figure*}[t!]
%{\includegraphics[width = 1.05\textwidth,trim={0cm 0.0cm 0cm 0cm },clip]{Fig4.png}} 
%\caption{\label{fig:surf} Surface streamlines plotted on (a) the washbasin wall, (b) on the ceiling, (c) on the floor and (d) on the wall adjacent to the door, along with contours of the x-component of wall shear stress.}
%\end{figure*}

By comparison, the recirculating regions in the corners are characterized by low velocity, with the white regions in the $x$-velocity plots indicating reversed flow. It is in the positive $x$-direction, opposite to the primary flow. A large region of reversed flow is present over the washbasin and it joins with the recirculating region formed due to the toilet seat to cover the entire length of the washroom. The streamline pattern in $y$ = 1 m cross-section shows imprint of the two vortices. The residence time of air in the recirculating regions is significantly higher than the primary flow, as the streamlines do not have a direct path of exiting the domain. It is in the range of 200 s to 800 s, while a small near-wall region on the top-right corner shows even higher residence time. This is because of a smaller secondary vortex formed in this corner, as indicated by the streamlines in the $y$ = 2 m plane. Similar corner vortex pattern, but to a smaller extent, is also present at lower height ($y$ = 1 and 1.5 m). The thin near-wall region with residence time of 1000 s (red color) is not expected to play a major role in the ventilation of infectious particles in the interior of the washroom.

% The three-dimensionality of the flowfield is more pronounced near the exhaust fan than the door, and as expected, the velocity magnitude is highest in the vicinity of the fan. 

\if{false}
The recirculation regions in the washroom are located next to walls and the ceiling. Surface streamlines plotted on the boundaries of the domain can be used to identify the extent of the recirculation zones. % This is based on the work of ?? on three-dimensional flow separation and attachment.
We can also identify limiting lines along with the surface streamlines converge or diverge, depending on whether the flow is locally leaving (separation line) or impinging on (reattachment line) the surface. 
In addition, the magnitude and sign of the wall shear stress is a good indicator of flow sepration and reattachment.

The washbasin wall in Fig.~\ref{fig:surf}a has two large regions of negative wall shear, along with streamlines directed backward towards the door in this region. These are the "foot-prints" of the recirculation region observed in the velocity vector plots in Fig.~\ref{fig:Fig2}, and we can identify the vertical extent of the reversed flow. All the streamlines in the washbasin region (both under and over the washbasin) coalesce into a limiting line that is parallel to the vertical edge of the room. This represents secondary separation, with a small secondary vortex at the corner. 
%%%
The other region of negative shear stress corresponds to the vortex formed adjacent to the toilet seat, and it is bounded by two limiting streamlines identified in the figure.

The opposite wall shown in Fig.~\ref{fig:surf}d has a large region of positive wall shear stress, which corresponds to the primary flow entering the washroom through the door. There is a small vortex being the door (blue region), and another small vortex at the other end of the wall. Otherwise, the air flow along this entire wall is directed towards the exit location.
%%%
An imprint of the primary flow is also visible on the floor as the red region of positive wall shear in Fig.~\ref{fig:surf}c. On the other hand, two negative shear stress regions are observed under the washbasin and next to the toilet seat. % The flow, however, is highly 
%%%
Finally, the surface streamlines on the ceiling show reversed flow over a large part and a limiting streamline near the top of the door.
\fi

% % % % % % % % % % % % % % % % % % % % % % % % % % % % % % % % % % % % % % % % 

\begin{figure}[t!]
\subfigure[$t$ = 50 s]{\includegraphics[width = 0.15\textwidth]{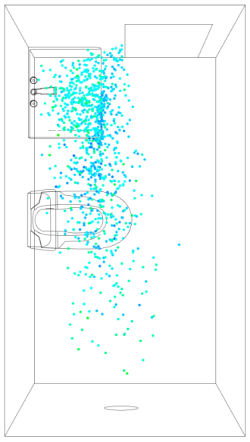}} 
\subfigure[$t$ = 100 s]{\includegraphics[width = 0.15\textwidth]{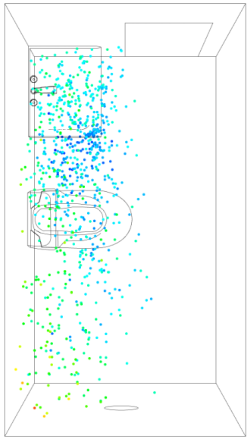}} 
\subfigure[$t$ = 200 s]{\includegraphics[width = 0.15\textwidth]{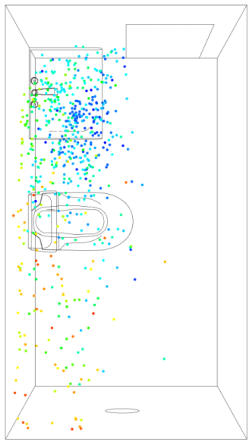}} 
\subfigure[$t$ = 300 s]{\includegraphics[width = 0.15\textwidth]{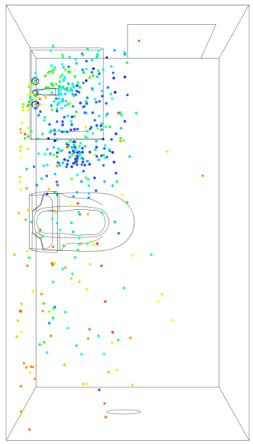}} 
\subfigure[$t$ = 400 s]{\includegraphics[width = 0.15\textwidth]{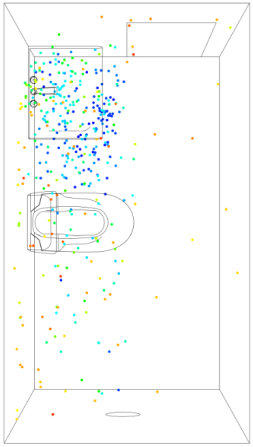}} 
\subfigure[$t$ = 500 s]{\includegraphics[width = 0.15\textwidth]{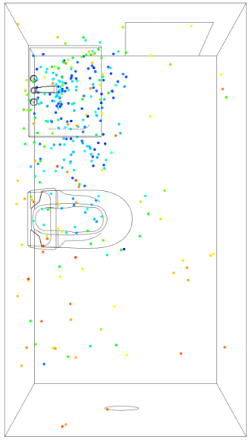}} 
{\includegraphics[width = 0.4\textwidth,trim={0cm 0.0cm 0cm 0cm },clip]{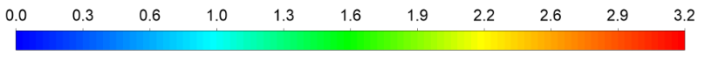}}
\caption{Spatial distribution of particle in the domain at different instants of time measured from the time of injection at the washbasin. The particles are colored based on their height, such that red particles are close to the ceiling, green particles are at the height of a person and blue particles are close to the floor. }
\label{basin_inj} 
\end{figure}

% % % % % % % % % % % % % % % % % % % % % % % % % % % % % % % % % % % % % % % % 

% % % % % % % % % % % % % % % % % % % % % % % % % % % % % % % % % % % % % % % % 

%\clearpage
\begin{figure}
\subfigure[Different zones]{\includegraphics[width = 0.45\textwidth, trim={0cm 0.cm 0cm 0cm },clip]{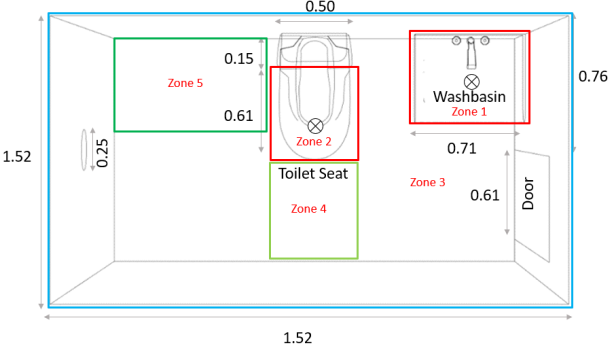}}
\subfigure[Particles in zones 1 and 2] {\includegraphics[width = 0.45\textwidth, trim={0cm 0.cm 0cm 0cm },clip]{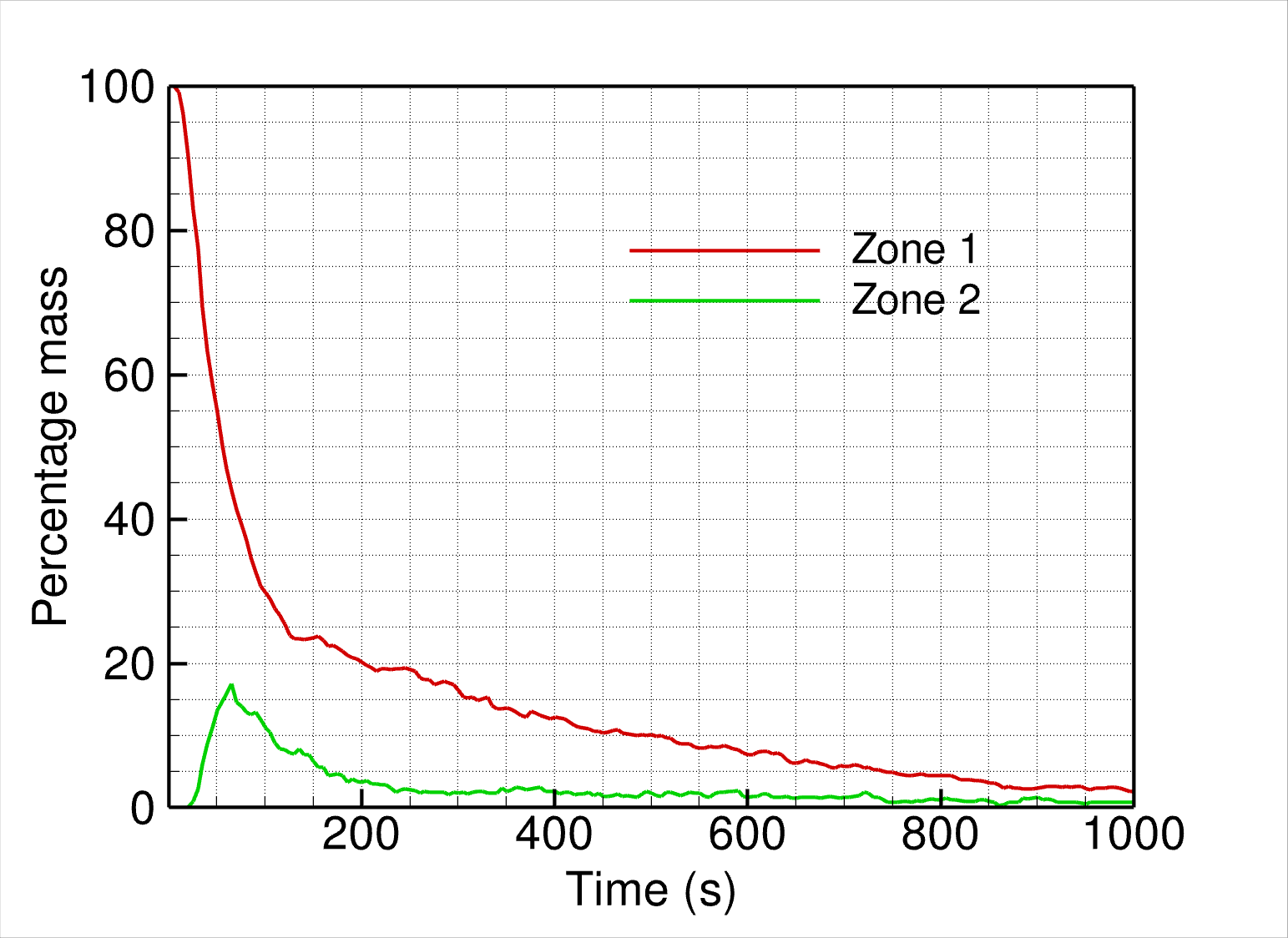}}
\subfigure[Particles in zones 3, 4 and 5] {\includegraphics[width = 0.45\textwidth, trim={0cm 0.cm 0cm 0cm },clip]{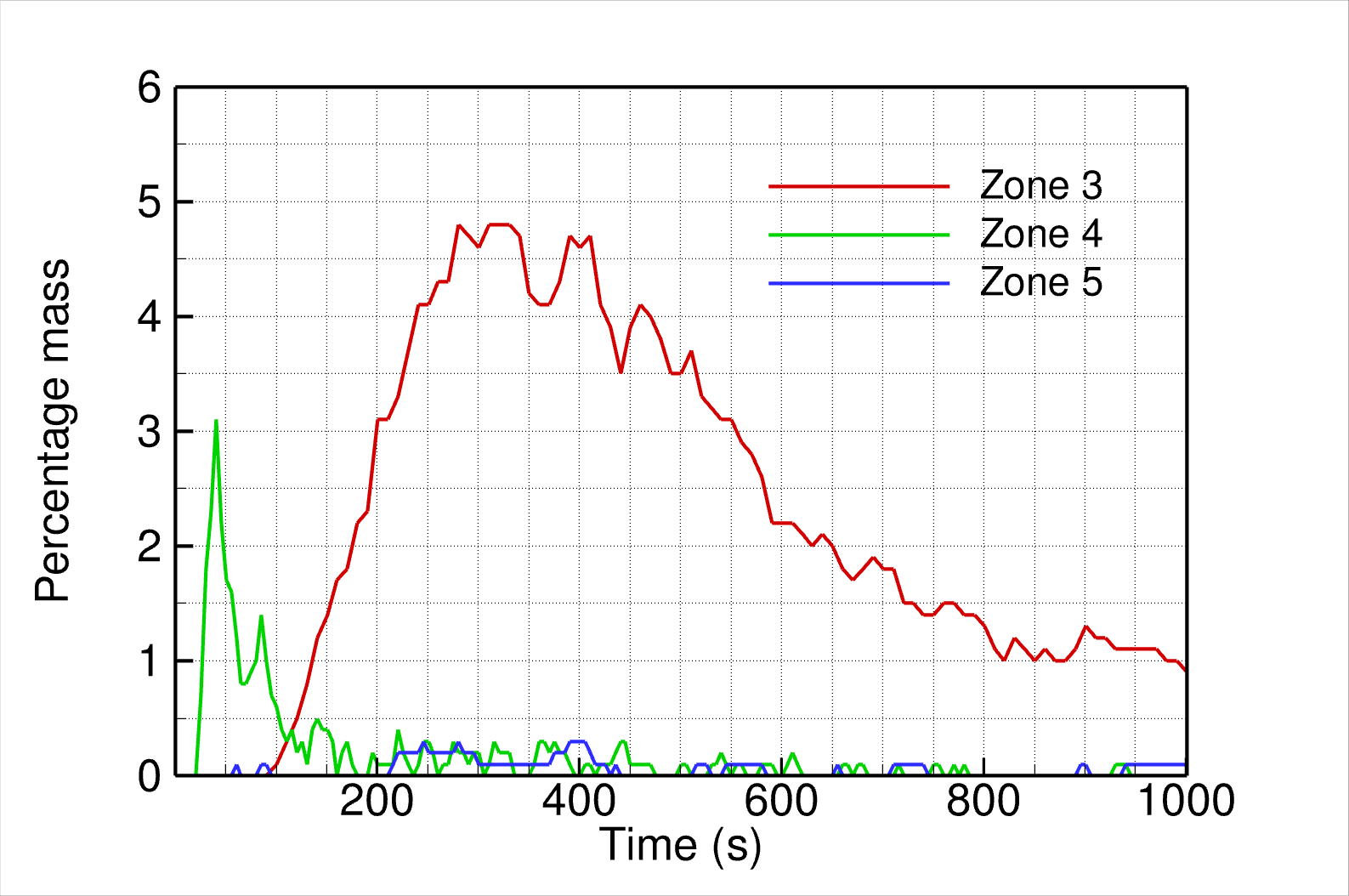}}
\caption{Time history of percentage particle mass, relative to the total mass of particles injected, in different regions of the washroom.}
\label{fig:zones}
\end{figure}

% % % % % % % % % % % % % % % % % % % % % % % % % % % % % % % % % % % % % % % % 

\subsection{Particle tracking results}

Particles are injected at the washbasin at $t$ = 0 and their locations are tracked over time. The particle tracking simulation details are given in section 2. 
Figure~\ref{basin_inj} shows the particle distribution in the washroom at different instants of time, from 50 s to 500 s. 
The particles disperse and spread in the domain, and are eventually ejected by the exhaust fan. Some particles, however, remain trapped at the washbasin (green color corresponding to a 1.3 < $y$ < 1.9 m), due to the recirculating flow in this region. Similarly, blue particles (for $y$ < 0.6 m) are trapped in the vortex formed under the washbasin.
At all times, there are more particles in the left half (recirculating flow) than the right half of the washroom (primary flow). 
Only exception is at large times ($t$ > 300 s), when particles tend to accumulate near the ceiling (red particles, $y$ > 2.5 m). The majority of the ceiling is in a recirculation region, with reversed flow, as shown in Fig. 2.

To study the distribution of the particles in the domain more quantitatively, we define several sub-domains or zones (shown in Fig. 6a). We choose two zones to cover the frequently-used areas of the washroom. Zone 1 is in the recirculating region at the washbasin, while zone 2 is over the toilet seat. %partly in the primary flow and partly in recirculating flow.
Zone 3 covers the entire ceiling, above the height of 2.5 m. Zone 4 is in the primary flow between the door and the exhaust fan, and zone 5 is at the corner next to the toilet seat, where we see a second recirculating vortex (in Fig. 2).

\begin{figure}
\subfigure[Particle distribution]{\includegraphics[width = 0.45\textwidth, trim={0cm 0.cm 0cm 0cm },clip]{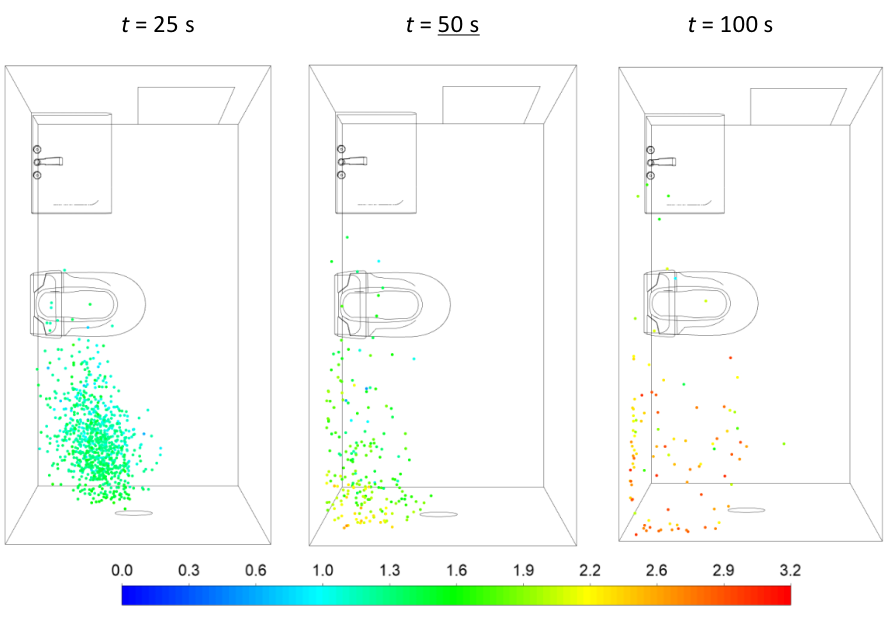}}
\subfigure[Percentage particle mass] {\includegraphics[width = 0.5\textwidth, trim={0.1cm 0.1cm 0.1cm 0.1cm },clip]{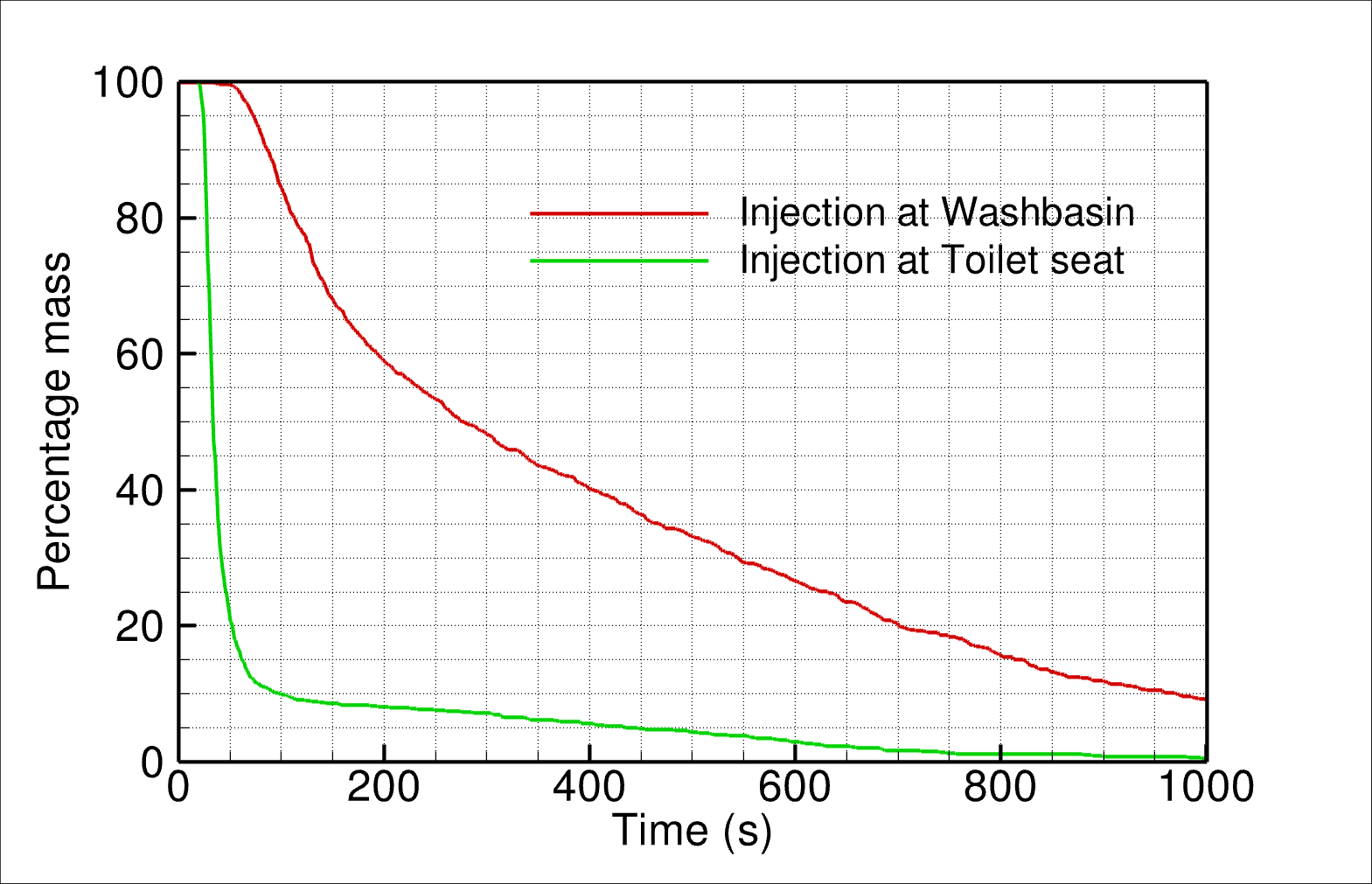}}
\caption{Particles injected at the toilet seat are tracked over time in terms of (a) their spatial distribution and height (shown by color), and (b) the mass of particles retained in the domain as a fraction of the initial mass injected at $t$ = 0.}
\label{fig:toilet_inj} 
\end{figure}

The percentage mass of particles in each zone is plotted as a function of time in Fig.~\ref{fig:toilet_inj}. It is once again relative to the total mass of particles injected in the domain at $t$ = 0. Note that all the particles are identical, and there is no change in their mass and size over time due to evaporation. Thus the percentage mass shown in Fig.~\ref{fig:toilet_inj} is equivalent to the number of particles in each zone relative to the total number injected initially.

As expected, the largest fraction of injected particles are in zone 1, all the particles are injected in this region at t = 0.
There is a rapid fall in the particle numbers in zone 1, as the particles disperse in the domain. At the same time, there is an increase in the particle numbers in zones 2, 3 and 4. 
Subsequently, the particle numbers decay in all the zones; the slowest decay is observed in zone 3 at the ceiling. This is in line with the fact that the particles get trapped in the large region of reversed flow at $y$ > 2.5 m. On the other hand, there are negligible particles in zone 5 in the corner next to the toilet seat.
Overall, at any given time, the number of particles in zone 1 is far greater than the other zones, indicating that the recirculation region formed over the washbasin can trap infectious particles for a long duration of time.

% % % % % % % % % % % % % % % % % % % % % % % % % % % % % % % % % % % % % % % % 

% % % % % % % % % % % % % % % % % % % % % % % % % % % % % % % % % % % % % % % % 

%\clearpage

Next, we inject particles over the toilet seat to simulate toilet flushing that generates a large amount of droplets.
The toilet seat is located partially in the primary air flow (see Fig. 2) and the particles injected here are rapidly carried away towards the exhaust fan (see Fig.~\ref{fig:toilet_inj}a). The majority of the particles are vented out of the washroom, while some get trapped at the ceiling (red particles with $y$ > 2.5 m). There are very few particles present in the washroom at $t$ = 100 s, as compared to the earlier case of washbasin injection (see Fig.~\ref{basin_inj}).
The total mass of particles in the domain is plotted as a function of time in Fig.~\ref{fig:toilet_inj}b, for both the injections. It shows a rapid decay for the toilet seat injection, driven by the primary flow time scale of $\tau_P$ = 50 s. There is a long tail after the initial drop, possibly because of particles trapped at the ceiling level.
%There is an initial period of 25 s, when there is no change in the particles
Only 10\% particles are left in the washroom after 100 s, compared to 90\% particles in the domain for the case of washbasin injection.
In the latter case, it takes more than 1000 s to bring down the particles numbers down to 10\% of the initial value. Thus, the injection in a recirculation zone results in ten times slower venting of the particles from the domain than injection in primary flow.

% =======================================

%\clearpage

% *******************  *******************

%\clearpage
\subsection{ Effect of fan CFM}

The air flow set up in the washroom is driven by the exhaust fan, and the volume flow rate of air through the fan is expected to play a crucial role. In the simulation, the volume flow rate is specified in terms of the velocity boundary condition prescribed at the fan outlet. We vary the exit velocity from 1 m/s to 3 m/s and study its effect on the air flow pattern and ventilation rate. Note that the fan is rated at 270 m$^3$/h, which translates to an exit velocity of 1.5 m/s. A lower exit velocity could be representative of normal wear and tear of the fan, so that it performs at a lower volume flow rate. A higher exit velocity, on the other hand, may be interpreted as increasing the speed of the fan or replacing it with a more powerful fan with a higher CFM.

\begin{figure}
\includegraphics[width = 0.5\textwidth, trim={0cm 0.cm 0cm 0cm },clip]{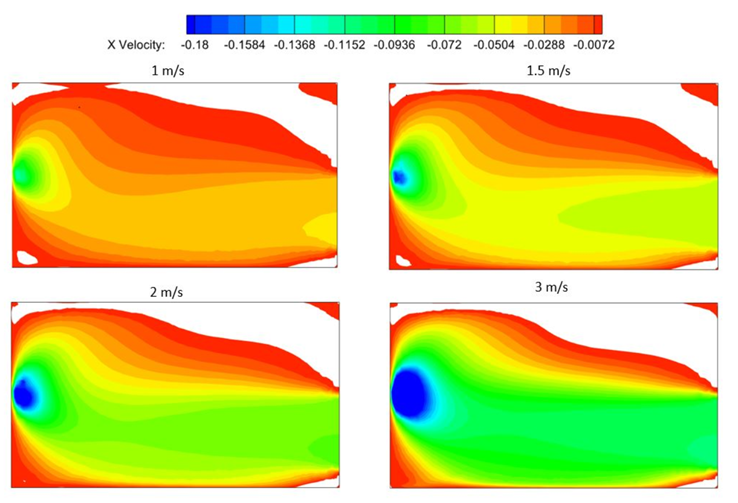}
\caption{\label{fig:cfm} Contours of $x$-component of velocity plotted on a horizontal plane at $y$ = 1.5 m for different values of the fan exit velocity. The white region corresponds to reversed flow (positive velocity), and it gives an indication of the size of the recirculation region at the four corners of the washroom.}
\end{figure}

Figure~\ref{fig:cfm} shows the simulation results for four cases, where the $x$-component of the velocity is used to study the effect of varying the exhaust fan CFM. The data is similar to that presented in Fig. 3, with high negative velocity in the primary flow between the door and the exhaust fan. The magnitude of flow velocity increases proportional to the exit velocity prescribed at the fan, but the qualitative flow pattern remains unaltered by changing the fan CFM. Specifically, the large reversed flow region formed on the washbasin, and marked by white color, is comparable in size and shape between the four solutions. 
%The same is true for the recirculation regions at the other corners of the 
Thus, changing the fan CFM is found to have negligible effect on the size of the reversed flow regions. Varying the CFM of the exhaust fan may not be able to eliminate the recirculation zones in the washroom.

% The primary effect of increasing the fan exit velocity is visible as higher magnitude air flow velocity near the fan location. We also see a slight increase in air flow velocity in the rest of the domain, particularly at the door entry location.
% These regions are identified as the white portions at the corners, because of a contour cut-off value of 0 m/s. 

% % % % % % % % % % % % % % % % % % % % % % % % % % % % % % % % % % % % % % % % 

\begin{figure}
{\includegraphics[width = 0.35\textwidth, trim={1cm 0cm 0cm 0cm },clip]{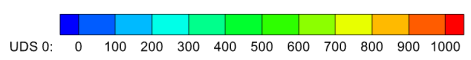}}
{\includegraphics[width = 0.45\textwidth, trim={0cm 1.5cm 1.cm 1.5cm },clip]{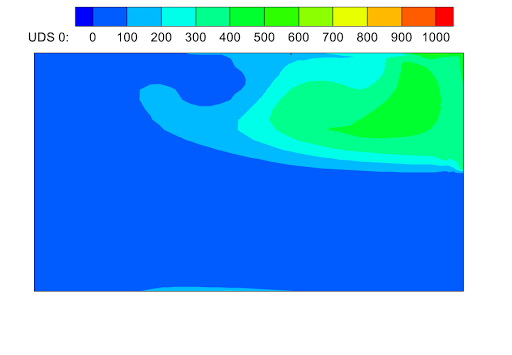}}
\caption{Air residence time (s) contours plotted in a horizontal plane at $y$ = 1.5 m for exhaust fan velocity of 3 m/s. The contour levels are identical to those in Fig,~2 for easy comparison with the residence time for exit velocity of 1.5 m/s.}
\label{res_cfm} 
\end{figure}
% % % % % % % % % % % % % % % % % % % % % % % % % % % % % % % % % % % % % % % % 

An increase in flow velocity in the domain, for a higher exhaust fan CFM, decreases the air residence time and vents the used air in the washroom more quickly. 
Fig.~\ref{res_cfm} plots the distribution of fluid residence time in a horizontal section at $y$ = 1 m, for fan velocity $V_{fan}$ = 3 m/s and it can be directly compared with the corresponding plot in Fig.~\ref{flowfield} for $V_{fan}$ = 1.5 m/s.
The two plots are qualitatively similar, with a higher residence time in the recirculating region, and lower values in the primary flow between the door and the exhaust fan.
The residence time varies between 100 to 400 s for the higher fan CFM, and it is lower than the corresponding values in Fig.~\ref{flowfield}.
We take a representative mean value for the recirculation time scale $\tau_R = $ 500 s and 250 s for $V_{fan} = $ 1.5 m/s and 3.0 m/s, respectively, and this is compared to the particle mass decay rate presented below.
The primary flow time-scale is also reduced by a factor of two, from $\tau_P = $ 50 s for the lower fan air velocity to 25 s for the higher exit velocity.
This can be interpreted as the time required for air to travel directly from the door to the exhaust fan in the well-ventilated parts of the washroom (primary flow regions marked in Fig.~\ref{vel_vect}), as mentioned in section III.A.

Next, we study the effect of the fan CFM on the rate of the decay of particle mass in the washroom.
If we assume the washroom as a fully-mixed reactor model, we can write the following equation for the particle concentration $C$ in the domain\cite{indoor_PNAS_2021}
$$ \frac{dC}{dt} = - \frac{Q}{\forall} C$$ 
where $Q$ is the volume flow rate of air through the fan and $\forall$ is the volume of the room. 
We have assumed no evaporation or sedimentation or deactivation. This will give a conservative estimate of the particle concentration in the room as a function of time. 
For an initial concentration $C_0$ at $t = 0$, 
%or equivalent initial number of particles $N_0$, 
we have 
$$ C = C_0 e^{-t/\tau_{ACH}} $$
where $\tau_{ACH} = \forall/Q$ is the ACH time scale based on the volumetric flow rate through the fan. 
For a fan exit velocity of 1.5 m/s, $\tau_{ACH}$ = 170 s, and it is reduced to 85 s for the 3.0 m/s fan exit velocity.

% / / / / / / / / / / / / / / / / / / / / / / / / / / / / / / / / / / / / / / / / / / / / / / / / / / / / / / / / / / / / / / / / / / / / / / / / / / / / / / / / % 

% <<<<<<<<<<<<<<<<<<<<<<<<<<<<<<<<<<<<<<<<< % 

\begin{figure}
\includegraphics[width = 0.53\textwidth, trim={0cm 0.cm 0cm 0cm },clip]{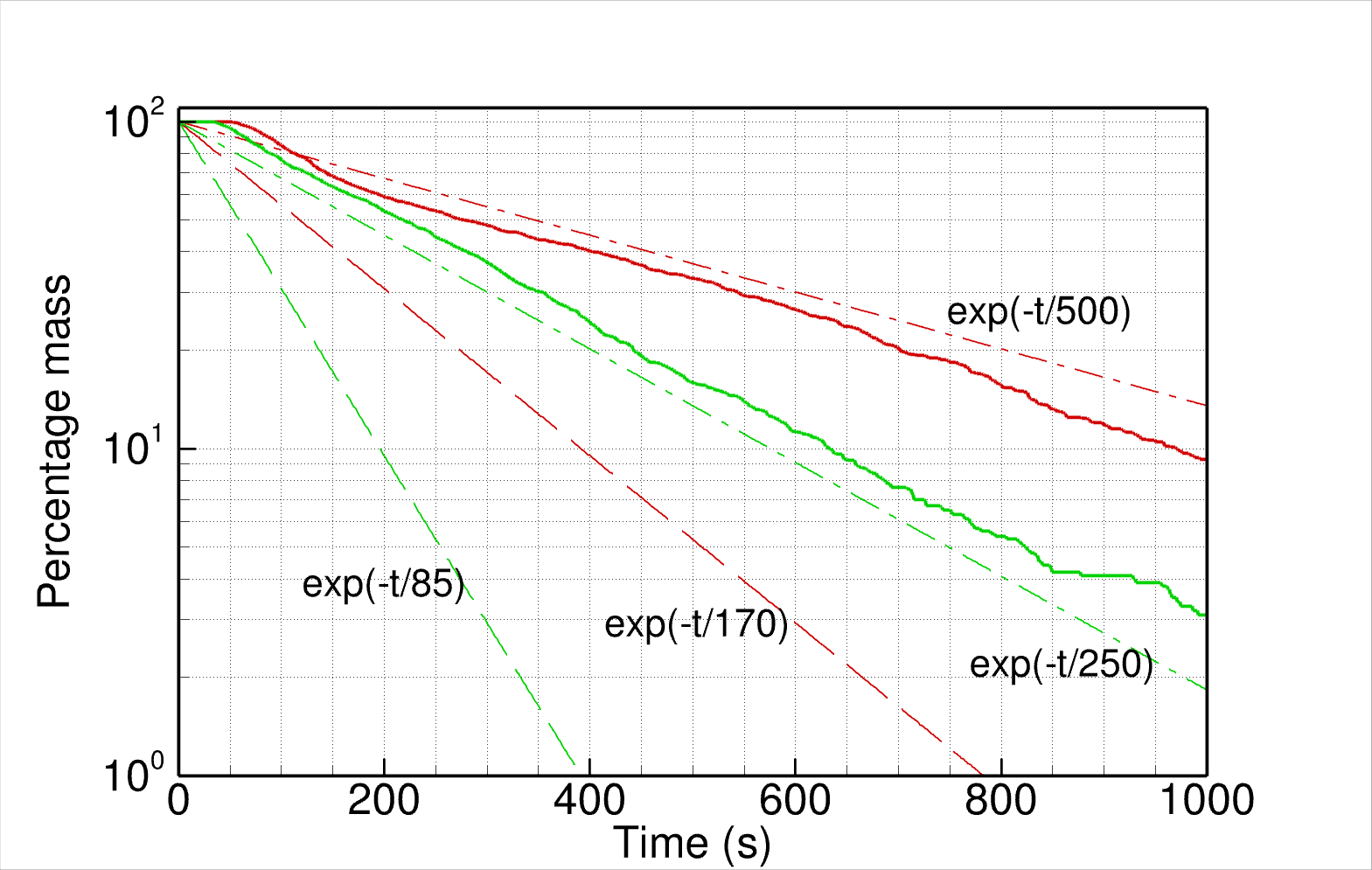}
\caption{\label{fig:cfm_decay} The rate of ventilation as indicated by the decrease in the particle mass remaining in the washroom, as a fraction of the total initial mass of particles injected at $t$ = 0. The data for fan exit velocity of 1.5 m/s (red) and 3.0 m/s (green) are compared with the respective theoretical curves.}
\label{fig:cfm_decay}
\end{figure}

% >>>>>>>>>>>>>>>>>>>>>>>>>>>>>>>>>>>>>>>>>> %
% ^^^^^^^^^^^^^^^^^^^^^^^^^^^^^^^^^^^^^^^^^^^^^^^^^^^^^^^^^^ % 

Figure \ref{fig:cfm_decay} compares the decay of particle mass in the domain with the ACH decay rate given above. The percentage of the initial particle mass injected (at $t=0$) is plotted for fan exit velocity of 1.5 m/s and 3.0 m/s. For a fixed volume of the domain, the percentage particle mass is equivalent to $C/C_0 \times 100$ and it is compared to the exponential term $e^{-t/\tau_{ACH}}$.
We have used a logarithmic scale to highlight the exponential trend of the fully-mixed reactor model.
%%%
It is clear from the figure that the particle mass does not follow the ACH decay rate that is frequently used to estimate the ventilation time required for a given volume flow rate.
Our results are significantly higher than that predicted by the fully-mixed reactor model,
implying that assuming the infectious aerosol to be uniformly distributed in the volume of air in the room can be grossly inadequate. 
This can result in significant under-estimation of the CFM required to achieve a prescribed ACH for a given room.

Interestingly, the CFD data follows an exponential decay with a characteristic time that is representative of the average recirculation time scale $\tau_R$ described above.
This is true for both the fan CFMs, and it points to the fact that the ventilation process in this scenario is dominated by the recirculating flow. 
The recirculation zones present in the washroom trap the air and the particles for a long time, potentially delaying the venting of infectious aerosol significantly.
Note that the recirculation time scale is about three times higher than the ACH time scale. Thus, at any time $t>0$, the number of particle in the domain is higher than the estimate of a fully-mixed air in the room by approximately a factor of $e^{2t/500}$. For example, at $t$ = 500 s, the ACH decay rate with $\tau_{ACH} = 170$ s would predict only 5\% particles remain in the domain, while the actual figure is close to 35\% in the corresponding CFD data with $V_{fan}$ = 1.5 m/s; an increase by a factor of about seven.

% Alternately, to reach a specified low level of particles in the room, the simulation data predicts a time duration that is ?? times longer than the ACH estimate.

% % % % % % % % % % % % % % % % % % % % % % % % % % % % % % % % % % % % % % % % 

\begin{figure}
\subfigure[zone 1]{
\includegraphics[width = 0.45\textwidth, trim={0.1cm 0.03cm 0.1cm 0.1cm },clip]{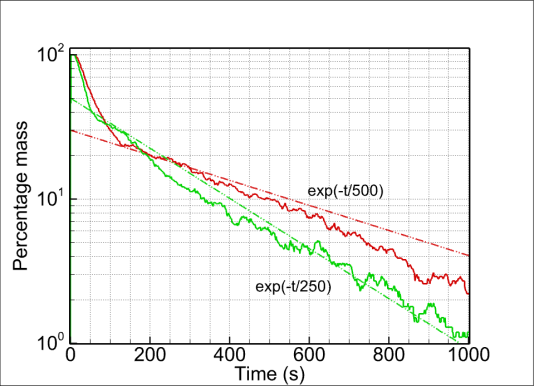}}
\subfigure[zone 2]{
\includegraphics[width = 0.45\textwidth, trim={0.1cm 0.03cm 0.1cm 0.1cm },clip]{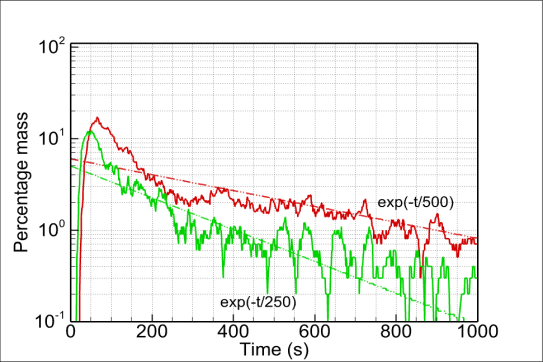}}
\caption{Comparison of the percentage particle mass in different zones for fan exit velocity of 1.5 m/s (red) and 3.0 m/s (green). The exponential decay with $\tau_R$ = 500 s and 250 s, respectively, are also shown for reference and the theoretical curves are offset on the vertical axis to account for the initial transient in each case.}
\label{fig:cfm_zone}
\end{figure}

% % % % % % % % % % % % % % % % % % % % % % % % % % % % % % % % % % % % % % % % 

The data presented in Fig.~\ref{fig:cfm_decay} is for injection in the washbasin for both the fan CFMs, as the washbasin can be an important source of droplets and aerosol. The other parameters of particle injection are kept constant between the two simulations.
As expected, %the higher air flow through the fan leads to lower particle mass at any given instant of time. Conversely, 
the time taken to reach a chosen threshold of mass fraction (say 30\%) is lower for the more powerful exhaust fan (350 s) compared to the lower CFM case (550 s).
In both the cases, the particle mass in the entire domain approximately follows the exponential decay governed by the respective recirculation time scale.
% Note that the exponential curves predict a doubling of ventilation time if the fan CFM is reduced by a factor of two.
%%
The same is true for the particle mass in the two frequently-used zones of the washroom, namely, zone 1 above the washbasin and zone 2 is over the toilet seat; see Fig.~\ref{fig:cfm_zone}.
The data computed for zone 1, using $V_{fan}$ = 1.5 and 3.0 m/s, show a rapid initial drop in particle mass fraction, followed by a gradual decay close to the respective recirculation time scale.
The two curves are qualitatively similar, except for higher slopes in the higher CFM case.

Zone 2 data in Fig.~\ref{fig:cfm_zone}b follow a similar trend, except for an initial transient that includes a buildup to a peak value, followed by a drop in particle mass and a relatively steady decay thereafter. Once again, the time scale of the initial transient as well as the subsequent decay are higher for the lower CFM simulation, resulting in a higher mass of particles left in the domain than the higher CFM case.
The CFD data for $V_{fan}$ = 1.5 m/s is close to the theoretical decay, while the higher $V_{fan}$ data is more noisy. This could possibly be caused by the low levels of particle mass in the domain (< 1\%) for $t$ > 500 s in this case. It may require initial injection of more number of particles to get better statistics. 
% Note that the theoretical exponential curves are offset on the vertical axis to account for the initial transient.

% There are minor difference in the initial transient, where zone 1 shows a rapid drop in particle numbers. This is becuase of a rapid dispersion and spread of the particles injected inside zone 1 to the rest of the domian. By comparison, there is an initial buildup of particles in zone 2, as the particles injected in the washbasin find their way to the toilet seat.
% Subsequently, both zones show a relatively steady decay, with a lower number of particles in the the higher CFM simulation than the lower CFM case.
%%
The important point to note is that the ventilation process is qualitatively similar irrespective of the fan CFM, indicating that the flow pattern and the flow processes are not drastically altered by increasing the fan exit velocity. The time scales are halved by doubling the fan CFM, and this doubles the rate of overall ventilation.

% The exponential decay is plotted in terms of the fraction of particles $N/N_0 = C/C_0$ as dashed straight lines on semi-log axes in Fig.~\ref{fig:cfm_decay}. Here, $N$ is the number of particles in the domiain and $N_0$ is its initial value.
%%%
% We also plot the exponential decay with a characteristic time $\tau_R$ of the recirculation zone, and this is found to match the decay rate observed in the simulation. 

% The particle numbers in zone 2 show an initial buildup, while particles injected in the washbasin find their way to the toilet seat, and then are ejected more rapidly by the primary flow in this region. The particle numbers fall below 5000 in about 100 s for both the fan exit velocities. The higher fan CFM, however, leads to a faster buildup and decay for t < 100 s. 

\subsection{Effect of boundary conditions and injection parameters}

\begin{figure}
\includegraphics[width = 0.5\textwidth, trim={0cm 0cm 0cm 0cm },clip]{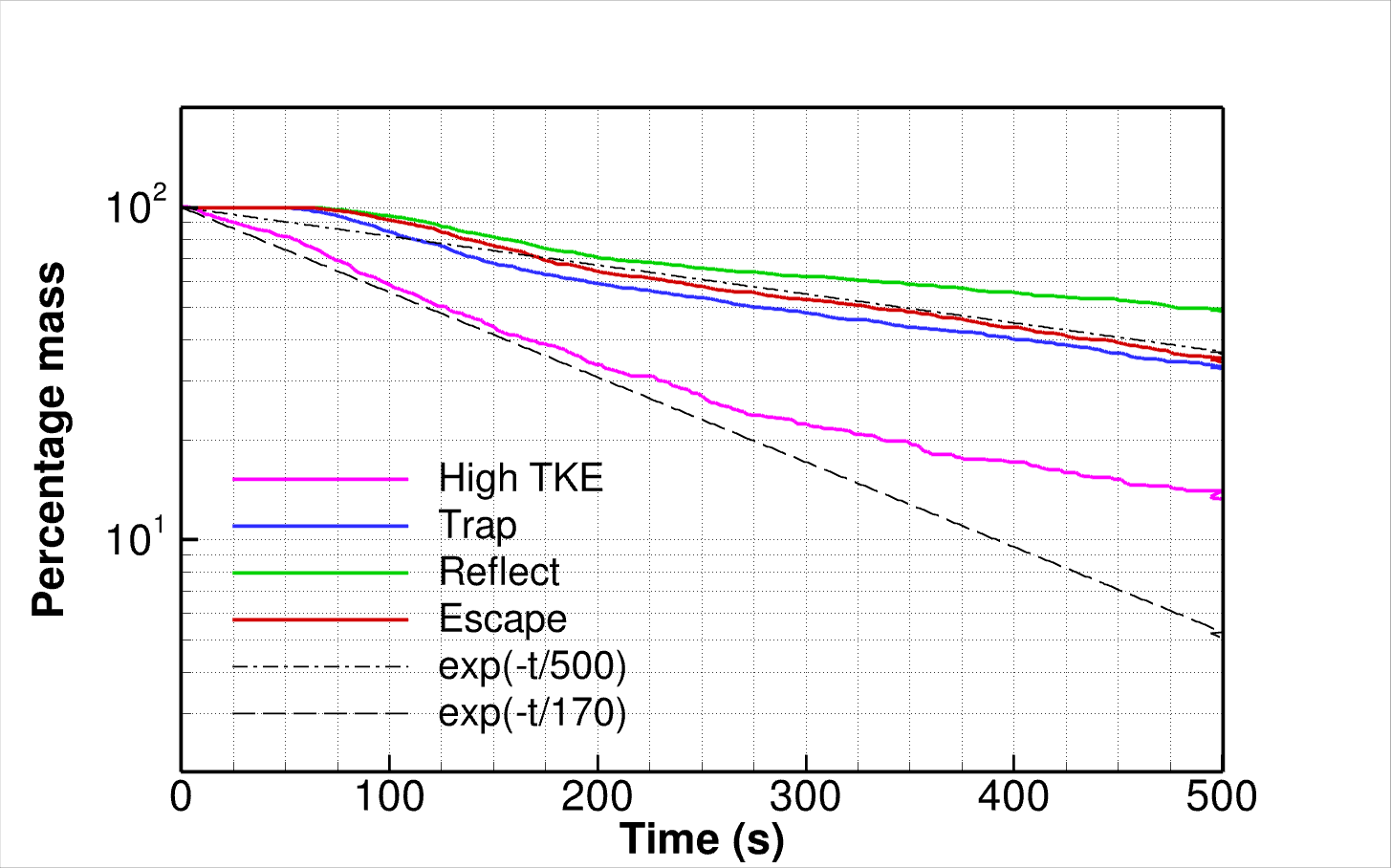}
\caption{\label{fig:turb_bc} The decay of particle mass in the domain obtained using different boundary conditions: higher turbulence level, particles reflect, escape or get trapped at the wall. These are compared with the exponential decay with $\tau_{ACH}$ (dashed line) and $t_R$ (dotted line).}
\label{decay_bc}
\end{figure}

We next study the sensitivity of the results presented above to the changes in the boundary conditions, both for the air flow simulation as well as the Lagrangian particle tracking.
Specifically, we vary the inlet value of the turbulence variables and whether the particles escape, reflect or get trapped at a solid boundary. All the cases presented in this section are for 1.5 m/s velocity at the exhaust fan, and particle injection at the washbasin. The results are compared in terms of the decay of percentage particle mass in the domain; see Fig.~\ref{decay_bc}.

We note that the earlier results are obtained by specifying a turbulence intensity of 5\% and a turbulent-to-dynamic viscosity ratio of 10. These give nominal values for the turbulent kinetic energy (TKE) and the turbulent dissipation rate at the inlet boundary. 
We compare the percentage particle mass presented earlier in Fig.~\ref{fig:cfm_decay} with that obtained using TKE and $\epsilon$ as 1 m$^2$/s$^2$ and 1 m$^2$/s$^3$, respectively. These extreme values can be interpreted as high turbulence levels generated at the door inlet due to disturbances in the air flow outside the washroom.
It is found that the particle mass fraction decreases more rapidly in the case with higher turbulence level. Interestingly, it follows the exponential decay with ACH time scale $\tau_{ACH}$ = 170 s for this case.
The percentage particle mass in the domain falls to 37\% at $t \simeq \tau_{ACH}$. At this point, the simulation results deviate from the theoretical ACH-decay curve, and asymptote to an exponential decay with a longer characteristic time.

% recirculation time scale $\tau_R$ for large time scales.

\begin{figure}
\subfigure[$t$ = 50 s]{\includegraphics[width = 0.15\textwidth, trim={0cm 0cm 0cm 0cm },clip]{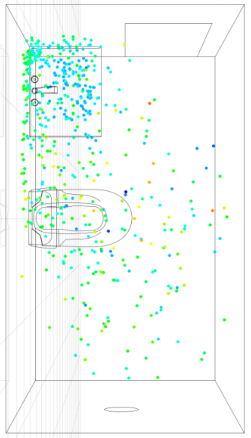}}
\subfigure[$t$ = 100 s] {\includegraphics[width = 0.15\textwidth, trim={0cm 0cm 0cm 0cm },clip]{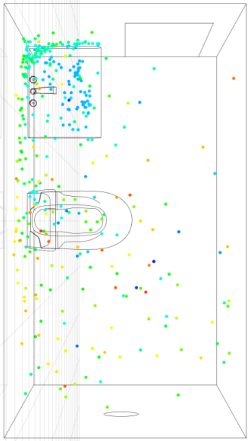}}
\subfigure[$t$ = 200 s] {\includegraphics[width = 0.15\textwidth, trim={0cm 0cm 0cm 0cm },clip]{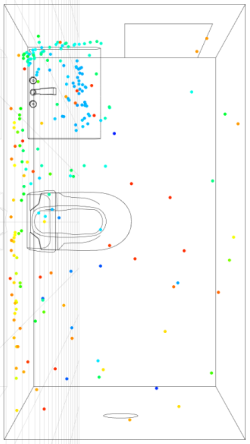}}
\caption{Spatial distribution of particles injected at the washbasin with varying time intervals, computed for the case with high turbulence level. The color legend is identical to Fig.~\ref{basin_inj} for direct comparison.}
\label{turb_inj} 
\end{figure}

The primary effect of varying the turbulence level appears in the rate at which the particles disperse out of the recirculation zone, where they are injected. In the high TKE simulation, the particles spread faster and farther in the room, and it is closer to a well-mixed reactor model of ventilation (Fig.~\ref{turb_inj}). See, for example, the spatial distribution of particles at $t$ = 100 s in comparison to the corresponding time instant in Fig.~\ref{basin_inj}b.
The decay rate of particle mass in the domain (at $\tau_{ACH}$) supports this observation. A fraction of particles remain trapped in the recirculation zone at the washbasin, and that gives the long tail of the decay curve. The slope of the high-TKE decay curve appears to be parallel to the earlier case, with characteristic time of $\tau_R = $ 500 s (see Fig.~\ref{decay_bc}).
The extent of recirculating flow is also altered by the turbulence level in the flow. A higher TKE gives a smaller recirculation zone at the washbasin compared to that shown in Fig.~\ref{flowfield} for the lower TKE calculation.
A smaller recirculation zone retains a lower number of particles, and this adds to the difference in the observed particle decay time history between the two cases.

The above results indicate that the ventilation rate can be enhanced by increasing the turbulence level in the air flow. This can be achieved by placing additional fans (for example, ceiling fans) to increase the mixing in the room. It can also reduce or disrupt recirculation zones, thereby aiding the exit of particles from the domain.
Depending on the level of TKE, the particle decay rate will vary between the timescales $\tau_{ACH}$ and $\tau_R$. A higher turbulence level gets the curve closer to the lower bound, given by the fully-mixed reactor model.
It is still possible that recirculation zones retain a fraction of the particles and a person placed in such regions is exposed to infection for a much longer time than the rest of the room.
A detailed mapping of the room in terms of flow residence time can be used to identify such potentially dangerous locations.
% In the cases without adequate mixing of air in the room, the ventilation rate is significantly lower than the ACH estimate, and retrofit designs based on the fully mixed reactor model can be grossly inadequate.

Figure~\ref{decay_bc} also shows that the changes in particle decay rate caused by varying the particle boundary condition at solid walls is relatively small. 
Comparable results are obtained when the particles are assumed to escape from the air flow at the boundary and when they are trapped on the bounding surface. Both these curves follow the $\tau_R$ exponential decay rate.
By comparison, higher particles mass is predicted when the particles reflect from a solid wall and re-enter the flow domain. 
The real scenario is expected to be in between the values predicted by the different particle boundary conditions, as a fraction of particles may reflect from a wall, based on their velocity and impact angle.
A conservative upper bound can be obtained when all particles reflect from the solid surfaces.

%\if{false}
\begin{figure}
\includegraphics[width = 0.5\textwidth, trim={0.1cm 0.1cm 0.1cm 0.1cm },clip]{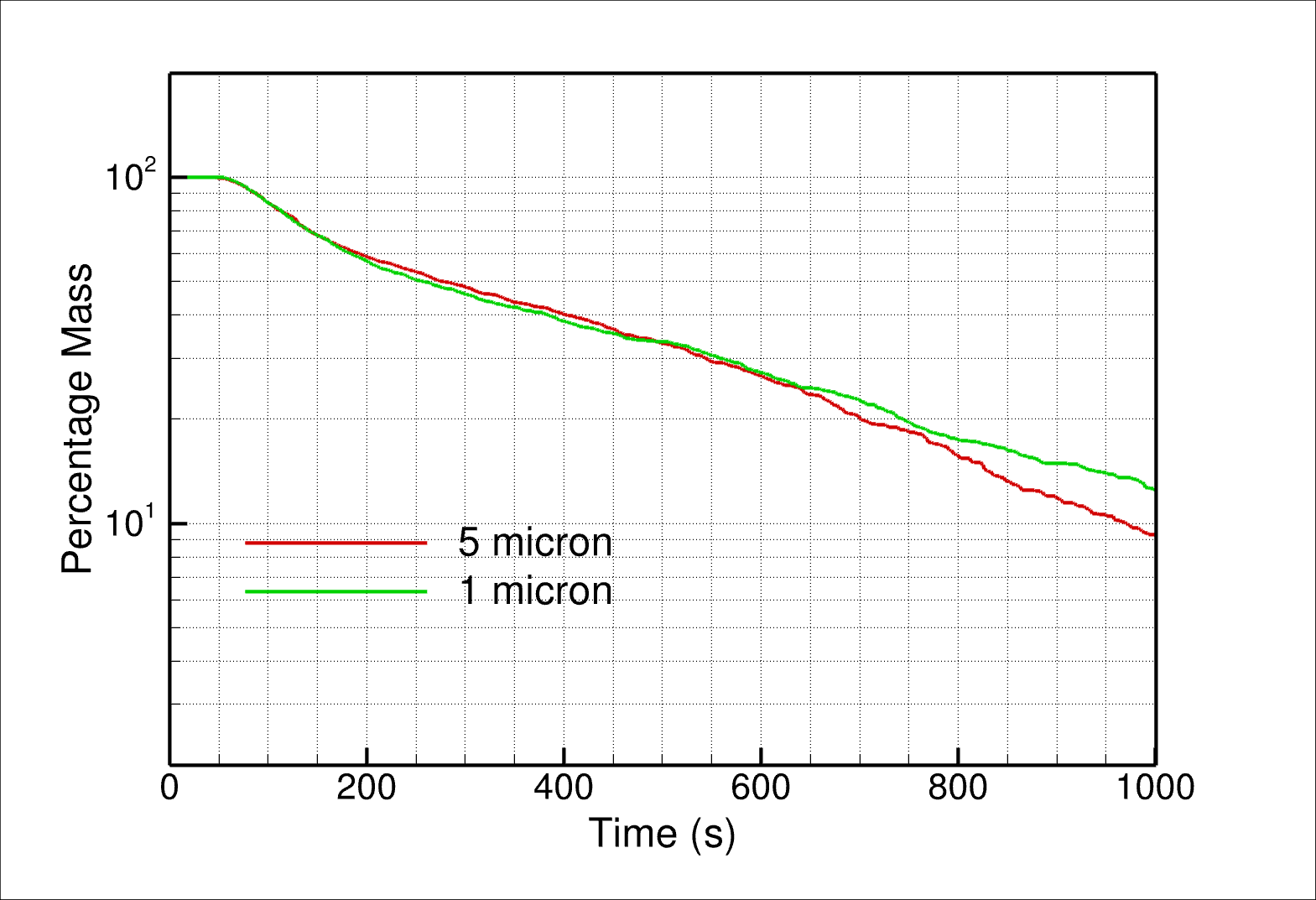}
\caption{\label{fig:turb_bc} The effect of particle diameter on the decay of particle mass in the domain. % These are compared with the exponential decay with $\tau_{ACH}$ (dashed line) and $t_R$ (dotted line).
}
\label{decay_dia}
\end{figure}
%\fi 

Finally, we study the effect of particle diameter and injection height on the ventilation results.
All other parameters are held constant: $V_{fan}$ = 1.5 m/s, 5\% turbulence intensity and escape boundary condition for particles at solid walls.
Increasing the particle diameter from 1$\mu$ (used in all the above simulations) to 5$\mu$ causes minimal changes in the time history of the particle mass in the domain.
The 5$\mu$ results overlap with the 1$\mu$ data 
(see Fig.~\ref{decay_dia})
%(shown with red line in Fig.~\ref{fig:cfm_decay}) 
up to 650 s, beyond which the larger particle simulation has a lower particle mass in the domain. 
This could be because of a higher gravitational effect, which accumulates over time and is noticeable towards the end of the simulation.
The maximum difference between the two diameter simulations is found to be about 10\% at $t$ = 1000 s.
%%%

The difference between the results obtained by varying the droplet diameter could also be interpreted as a margin of error caused by neglecting evaporation.
The larger droplets (5$\mu$) would evaporate to a smaller size (1$\mu$) during the course of the simulation, %such that the effects of gravity and evaporation would compensate each other.
and thus the results presented up to 1000 s are expected to be valid (within 10\%), even with evaporation, for small particle diameters up to 5$\mu$. 
Droplets of significantly higher diameter, in the range of 100$\mu$, will fall to the ground much sooner.
Sedimentation of large droplets and surface contamination are beyond the scope of the current study, and will be taken up in the near future.

\if{false}
\begin{figure}
{\includegraphics[width = 0.45\textwidth, trim={0cm 0cm 0cm 0cm },clip]{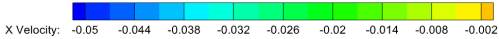}}
\subfigure[$x$-velocity]{\includegraphics[width = 0.45\textwidth, trim={0cm 0cm 0cm 0cm },clip]{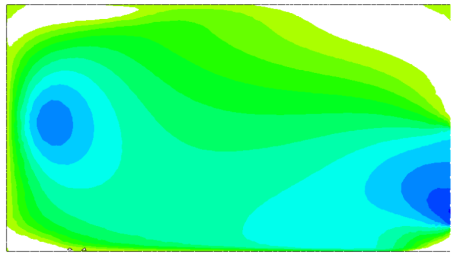}}
{\includegraphics[width = 0.45\textwidth, trim={0cm 0cm 0cm 0cm },clip]{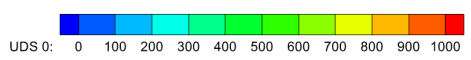}}
\subfigure[air residence time] {\includegraphics[width = 0.45\textwidth, trim={0cm 0cm 0cm 0cm },clip]{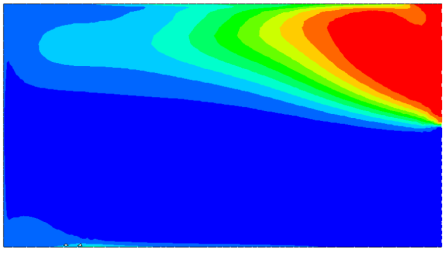}}
\caption{}
\label{fig:turb_cf} 
\end{figure}
\fi

\if{false}
\section{Discussion}

recommendations for ventilation

We note that the air flow pattern in an enclosed space is dependent on its geometry and layout. Changes in door orientation and windows can make a substantial difference to the primary and secondary flow regions. This can be used effectively to eliminate recirculating air from highly-used areas of a room.
% aid air flow management, as shown in this work.
% reduce the residence time of air, especially in the heavily-used areas of a room.
For example, in rooms with multiple windows, opening and closing of different windows can tailor the air flow pattern. It can direct incoming fresh air towards areas that are used by multiple people. This can significantly aid in air flow management to mitigate risk of air-borne infection.
The incense smoke technique can be easily extended to classrooms, restaurants, shared office spaces and other indoor environment.
The simplicity of the experimental set up make it readily amenable to field studies.

%This provides an easy and readily available mode to maximize the region of primary and minimize or eliminate dead zones. secondary flow .

$ $ \hrule $ $

\fi

\section{Conclusion}

In this work, we use computational fluid dynamics of air flow and Lagrangian particle tracking to study the ventilation of a public washroom. We consider a single-person washroom that is used by multiple people one after another. The objective is to replace the air in the washroom after every use, so as to vent out any infectious aerosol generated by a user. This is achieved by an exhaust fan that throws air out of the washroom and by keeping the door open between to consecutive usages.

The exhaust fan sets up a primary flow between the door and the exit location, so that the air in this region, along with any infectious particles, are quickly ejected from the washroom within 50 s.
There are, however, recirculating flow in the corners and at the ceiling level, which can harbor particles for much longer duration.
The frequently-used washbasin located next to the door, is one such region, and it is of particular interest.
Particles injected in the washbasin disperse in the domain and get ejected by the fan, but the rate of ejection is governed by the residence time of air in the recirculating flow. The recirculation time scale of 500 s is about ten-times higher than the primary flow time scale. It is also substantially higher than the ACH time scale of 170 s, based on the fully-mixed reactor model of the washroom.

Increasing the fan CFM can significantly reduce the ventilation time required to achieve a desired low level of particle concentration in the room. Also, increasing the turbulence level, via additional fans placed inside the washroom, can bring the ventilation rate closer to that of a fully-mixed reactor model. However, there can still be pockets of trapped air and higher particle concentration in recirculation zones. It is recommended that such regions are identified in a given room, and frequently-used equipment and people are not placed at such locations.

% In this work, we study the ventilation of a single-person public washroom that is used by multiple people one after another. The objective is to replace the air in the washroom after every use, so as to vent out any infectious aerosol generated by a user. This is achieved by an exhaust fan and by keeping the door open between to consecutive usages of the washroom.

\section*{Acknowledgement}
The authors acknowledge the fruitful discussions with Dr. Guruswamy Kumaraswamy of IIT Bombay during the course of this work.

\section*{Data Availability Statement}
The data that supports the findings of this study are available within the article.

%\clearpage

\section*{References}
\nocite{*}
\bibliography{aipsamp}% Produces the bibliography via BibTeX.

\end{document}